\documentstyle[12pt]{article}

\newcommand{\ben}{\begin{enumerate}}
\newcommand{\een}{\end{enumerate}}
\newcommand{\be}{\begin{equation}}
\newcommand{\ee}{\end{equation}}
\newcommand{\bse}{\begin{subequation}}
\newcommand{\ese}{\end{subequation}}
\newcommand{\bea}{\begin{eqnarray}}
\newcommand{\eea}{\end{eqnarray}}
\newcommand{\bc}{\begin{center}}
\newcommand{\ec}{\end{center}}

\def\ket#1{|#1\rangle}
\def\bra#1{\langle#1|}
\def\Zbf{{\bf Z}}

\begin{document}	

\begin{flushright}
NIKHEF/2000-001\\
Januari 2000
\end{flushright}

\vspace{15mm}
\begin{center}
{\Large\bf\sc Open strings, simple currents and fixed points}
\end{center}
\vspace{2cm}
\begin{center}
{\large A.N. Schellekens}\\
\vspace{15mm}
{\it NIKHEF Theory Group\\
P.O. Box 41882, 1009 DB Amsterdam, The Netherlands} \\
\end{center}

\vspace{2cm}

\begin{abstract}
Some applications of simple current techniques and fixed point resolution
to theories of open strings are discussed. 
In addition to a brief review of work presented in two recent papers
with L. Huiszoon and N. Sousa, some new results concerning uniqueness of
crosscap coefficients are presented, as well as a strange sum rule for
the modular matrix implied by the existence of crosscaps.

\end{abstract}
\section{Introduction}

Open strings have enjoyed a rather varied amount of interest during
the past three decades of string theory. In the early days
 open strings were used to describe mesons,
with quarks attached to the endpoints. Indeed, the Chan-Paton labels
still used today date back to as early as 1969 \cite{Paton:1969je}. This provided the first
method for obtaining gauge groups in string theory. The possible
gauge groups were classified much later \cite{Marcus:1982fr}, in a period when string
theory in general had fallen into decline. In 1983 Alvarez-Gaum\'e and
Witten \cite{Alvarez-Gaume:1984ig} showed that open superstrings (type-I)
were plagued by chiral anomalies
for any gauge group, whereas closed superstrings (type-II) were
automatically anomaly free. Although this looked like a fatal blow,
there was a brief revival after Green and Schwarz \cite{Green:1984sg}
 found a novel
mechanism to cancel the anomaly for the gauge group $SO(32)$. But
within months the heterotic string was discovered \cite{Gross:1985dd}.
 This theory 
could in addition to $SO(32)$ have the gauge group $E_8 \times E_8$ \cite{Thierry-Mieg:1985jf},
which at first sight seemed phenomenologically more attractive. 
During the subsequent ten years open strings were almost completely
neglected in favor of heterotic strings, which went through a phase
of rapid development. Apart from being phenomenologically disfavored,
open strings looked ugly and complicated:
their description requires world-sheets
with boundaries and crosscaps, and to obtain finite one-loop diagrams
one has to cancel tadpoles by hand. 

All this changed drastically in 1995, for several reasons. First of all
open strings were found to be part of the duality 
picture, and in particular in ten dimensions the strong coupling
limit of the type-I string was conjectured to be the heterotic string \cite{Polchinski:1996df}. 
Secondly, the discovery of D-branes swept out by the endpoints of
open strings made boundaries more respectable \cite{Polchinski:1994fq}. Furthermore in some
cases tadpole cancellation was found to be equivalent to charge
cancellation between D-branes and orientifold planes \cite{Polchinski:1995mt}, which sounds
somewhat less {\it ad hoc}. Furthermore it was pointed out that
the relation between the gauge and gravitational coupling is different
in open string theories and in closed ones,  which makes it possible to
separate the unification scale and the string scale \cite{Witten:1996mz}. 
This has opened new avenues in string phenomenology, involving
open strings (see {\it e.g.} \cite{Antoniadis:1998ig}).
These developments make
it clear that open strings must be considered seriously again.

\section{Closed strings, CFT and modular invariance}

During the period 1984-1994 there has been a lot of progress in 
the description of lower-dimensional closed strings 
in terms of conformal field 
theory (here and in the following ``closed" is an abbreviation
for ``closed and oriented").
This had led to a very economical formalism based on a few
simple, algebraic constraints, from which very general theorems
can be derived. It would be nice to have a similar description of open
strings. 
The algebraic constraints in the closed
string case are (essentially)
Lorentz invariance in the number of dimensions
one considers, (super)conformal invariance and modular invariance.  
If one builds the internal (``compactified") part of the theory
out of some (super)CFT, the constraint of modular invariance is
that the integer matrix $Z$ appearing in the one-loop partition function
\be
 P(\tau,\bar\tau) = \sum_{ij}\chi_i(\bar\tau) Z_{ij}\chi_j(\tau)
\ee
must commute with the generators $S$ and $T$ of the modular group
of the torus. Here $\chi_i(\tau)$ are the Virasoro characters of
the internal CFT (which may be combined with the space-time CFT in 
a more intricate way than suggested here, but this is easy to take
into account). Modular invariance is a simple and powerful constraint,
from which one can for example derive the Green-Schwarz factorization
of chiral anomalies \cite{ScW} or prove the existence of fractional
charges in a large class of heterotic string theories \cite{SchE}. 

Unfortunately all this is limited to perturbative, closed string
theories.  
There may be non-perturbative states in the spectrum 
of closed strings that are not controlled
by the modular invariant partition function. Nevertheless modular
invariance is, in my opinion, too nice a principle to simple give up.
I would hope to find some sort of generalization, a principle
that governs the presence or absence of states in string theory, 
M-theory or whatever string theory generalizes to. 
Such a principle would justify the term ``theory of everything".
Although this expression has fallen out of favor because it is
usually maliciously misinterpreted by adversaries of string theory
in particular and science in geneneral, it is justified in 
the following precise sense. 
In field-theoretic descriptions of our world it
is always possible to add some new particles to a
successful theory (and in particular to the standard model). There
are few theoretical constraints, but if one makes the new particles
sufficiently massive, unstable and weakly coupled, it is not hard to evade
all experimental and cosmological constraints.  This implies that
one can never claim to have arrived at a complete description of
all physics in our universe, since experimental and cosmological
constraints are always limited to some subset of any
relevant parameter space. This seems like an inevitable fact: one
can never know more than one has measured. However, string theory
provides a potential way out. Adding extra particles to a given
string theory makes it inconsistent. The only thing that one may
do is remove some states, and add others in their place. This is
perhaps best known in the example of orbifold constructions, where
one removes states from the spectrum that are not invariant under
a certain symmetry, and replaces them by ``twisted states". This
is not limited to orbifold constructions, but is in fact a general
property of modular invariant, perturbative closed strings. If it
generalizes beyond perturbative closed strings it would in principle
be possible to make a unique choice among the huge amount of string
vacua, based on a finite number of experimental results. If further
experiments find additional particles not predicted by 
this particular string theory,
then they can only be accomodated at the expense of some particles
that weregeneral found before. In other words, any further experiments would
rule out string theory as a whole, or in still other words, we
would have a very strong prediction for the existence or non-existence
of any other particle, no matter how massive or weakly coupled, in
our universe. This would certainly deserve the name ``theory of
everything". \input epsf\vskip .7truecm
\let\picnaturalsize=N
\def\picsize{3.5in}
\def\picfilename{SFcomp.eps}
\ifx\nopictures Y\else{\ifx\epsfloaded Y\else  \fi
\let\epsfloaded=Y
\centerline{\ifx\picnaturalsize N\epsfxsize \picsize\fi \epsfbox{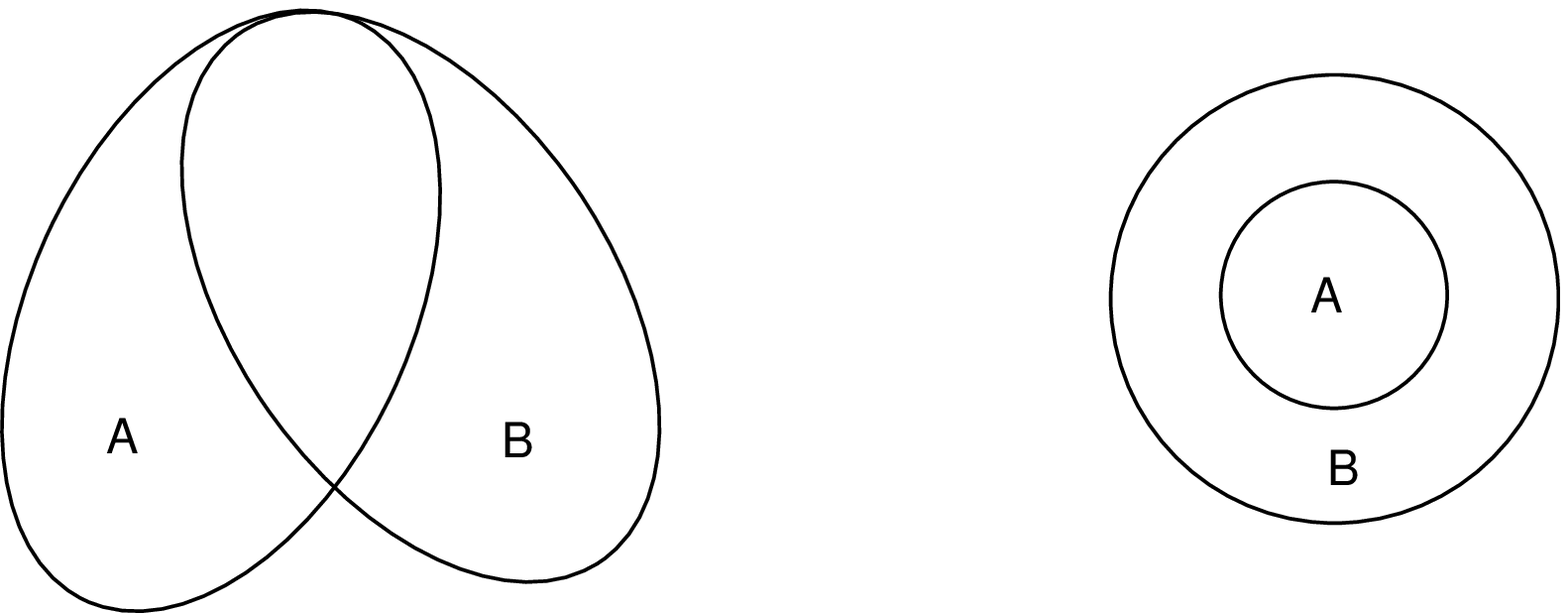}}}\fi
\vskip .5truecm
The argument given above is summarized in the above two pictures,
with on the left the string theory way of going from theory A to
a different theory B, and on the right the field theory way. 

At present these pictures are only a caricature of reality. There
is little hope of finding the right string ``vacuum" without
additional information, even if the first picture is the right one.
But in addition non-perturbative effects, and in particular
open and unoriented strings (which are non-perturbative from the point of view
of closed strings) pose a serious challenge to this picture.

Open and unoriented strings are usually constructed by starting
from a consistent closed, oriented theory. To the one-loop closed string
amplitude, the torus, one adds an unoriented closed string diagram,
the Klein bottle. This acts as a projection, removing certain
states from the spectrum. Furthermore one adds open string diagrams,
which at the one-loop level are the annulus and the Moebius strip. 
These diagrams come with a free parameter, the Chan-Paton multiplicity,
for each boundary. 
Until this point this looks reminiscent of the orbifold procedure,
with the Klein bottle projection playing the r\^ole of the removal
of non-invariant states, and with open strings 
playing the r\^ole of
twisted sectors. However, the presence of ``twisted" sectors is
in this case not governed by modular invariance, but by a 
different principle, the cancellation of massless tadpoles that 
lead to infinities. Unlike modular invariance this is
a target space criterion, and hence one loses the nice feature
of closed oriented theories  that a consistent world sheet theory
is sufficient to get a consistent target space theory. Worse yet,
there exist unoriented string theories for which the tadpole cancellation 
conditions require all Chan-Paton labels to vanish \cite{Angelantonj:1996mw}\cite{DaPa}.
 This means
that there are no open string states at all, even though the Klein
bottle still acts as a projection. Hence in this case the
states of the closed, unoriented  theory are a subset of those
of the closed oriented theory, as in the second picture above. Note
that this can never happen in modular invariant partition functions.
It is easy to show that if $Z_{ij}$ yields a modular invariant,
then any matrix $0 \leq \hat Z_{ij} \leq Z_{ij}$ can only be
modular invariant if $\hat Z=Z$ (this follows from 
$Z_{00}=\hat Z_{00}=1$ and $S_{0i} > 0$).

This implies that the first picture does not hold for perturbative
states. It might still be saved by non-perturbative states, if
the unoriented theory has non-perturbative states not present
in the oriented theory. On the other hand it is possible that 
the first picture has to given up, and that one has to allow for
a discrete set of consistent truncations of certain string theories. 
This does not necessarily invalidate the discussion given above. 
Most people would probably agree with the statement that in string
theory and its generalizations the possibilities for adding or removing
states are severely limited by consistency requirements. It would be 
nice to make that more precise.

\section{Simple currents and fixed points}

In search of a principle that governs the presence of states
in a string theory, I now turn to something much more down-to-earth,
namely a tool that plays an essential r\^ole in dealing
with modular invariance in the closed, oriented case: 
simple currents \cite{Schellekens:1989am}.
Simple currents are primary fields that upon fusion with any
other field yield just one field \cite{Schellekens:1989am}\cite{Intriligator:1990zw}. They can be used 
to build a non-diagonal partition function. If one has a closed
set of integral spin simple currents, these currents can extend the
chiral algebra, and one obtains a partition function
\be
 \sum_{Q(i)=0} N_i \big| \sum_{j \in \rm{Orbit}(i)} \chi_{j} \big|^2
\ee
Here $Q$ is a charge (or set of charges) defined for each current,
and the orbit is the set of distinct fields generated by
the set of currents acting on $i$. In general some currents may
fix $i$, and then the action of the currents covers the orbit
$N_i$ times. Fractional spin currents also generate modular invariants,
but they correspond to automorphisms of the fusion algebra, which
pair the left and right representations in an off-diagonal way. 

Simple currents are used for a variety of purposes in the 
construction of closed string theories, such as
\begin{itemize}
\item{Field identification in coset CFT's \cite{Schellekens:1990uf}\cite{Fuchs:1996tq}}
\item{World-sheet supersymmetry projections \cite{Schellekens:1990wx}}
\item{Space-time supersymmetry projections \cite{Schellekens:1990wx}} 
\item{D-type invariants \cite{Cappelli:1987xt}\cite{Bernard:1987xy}
\cite{Schellekens:1989am}\cite{Intriligator:1990zw}} 
\item{Inverse orbifolds (under conditions discussed in \cite{Fuchs:1999zi})} 
\end{itemize}
As we will see, they play a useful r\^ole in open string
constructions as well.

Another concept that comes back in open string theories
is the resolution of fixed points \cite{Schellekens:1990uf}\cite{Schellekens:1990xy}. 
The presence of the
multiplicities $N_i$ in partition functions implies usually
(but not always \cite{Fuchs:1996dd}), that the
corresponding terms are reducible representations of the
extended chiral algebra. To describe the modular properties of
the characters of the extended algebra 
(the orbit sums) 
one needs a set of matrices
that act on the resolved fixed points. There is such a matrix
$S^J_{ij}$ for any current that has fixed points, and it is
defined only on the fixed points $i$ and $j$ of $J$. In terms of
these matrices, and the original matrix $S\equiv S^0$, the
matrix $S$ of the extended theory takes the form \cite{Fuchs:1996dd}
\be\label{eq:sform}
S_{(i,\mu)(j,\nu)}=
{|{\cal G}| \over \sqrt{|{\cal S}_i||{\cal U}_i||{\cal S}_j||{\cal U}_j|}}
\sum_J \Psi^J_{\mu} S^J_{ij} (\Psi^J_{\nu})^* 
\ee
Here $\mu,\nu$ label the components into which the orbits of $i$ and $j$
are resolved, ${\cal G}$ is the simple current group that extends
the chiral algebra, and ${\cal S}_i$ (the stabilizer)
is the subgroup that fixes
$i$. The group ${\cal U}_i$ is a subgroup of ${\cal S}_a$ called
the {\it untwisted stabilizer}; for the precise definition see \cite{Fuchs:1996dd}.
The factors $\Psi$ are discrete group characters
of  ${\cal U}_i$, and $i$ is resolved into $|{\cal U}_i|$ components.

In general one can derive a list of properties that the matrices
$S^J$ should satisfy in order for $S_{(i,\mu)(j,\nu)}$ to be a proper
modular transformation matrix. In the case of WZW-models
one can find a natural set of matrices that satisfy all those
properties, and that are therefore obvious candidates for $S^J$ \cite{Schellekens:1990xy}\cite{Fuchs:1996dd}. 
This works as follows \cite{Fuchs:1996zr}. To each WZW-model belongs 
a Dynkin diagram, which is an extended
Dynkin diagram of an ordinary Lie algebra. 
Simple currents are related to symmetries of this extended Dynkin
diagram that move the extended root (with one exception for $E_8$
level 2 \cite{Fuchs:1991wb}). 
Given a Dynkin diagram symmetry one can define a folded
diagram in a fairly obvious way (for details see \cite{Fuchs:1996zr}), and
to that diagram one can associate a Cartan matrix. This defines
a new algebra, which we call the {\it orbit Lie algebra}. 
There is such an algebra for any simple current, and the 
modular transformation matrices of the characters of this orbit
Lie algebra are equal to the matrices $S^J$, up to a 
calculable phase. The orbit Lie algebras of simple currents
are affine Lie algebras, so that their modular transformation
matrices are calculable using the Kac-Peterson formula \cite{KaPe}. In one
case the orbit Lie algebra is a {\it twisted} affine Lie algebra,
but luckily this is precisely the only twisted affine Lie algebra
whose characters have good modular properties. 

The generalization of the foregoing to arbitrary CFT's is not 
completely understood yet. The concept of an
orbit Lie algebra seems restricted to WZW-models. But 
in any case most known rational CFT's are related to WZW-models
by the coset construction. Since field identification can be
described in terms of simple currents (with a few exceptions), formula
(\ref{eq:sform}) applies to those cases. One can apply formula 
(\ref{eq:sform}) to a
combination of field identification and any chiral algebra extension of
the coset theory, and read off the matrices $S^J$ of the coset theory. 
In \cite{Schellekens:1999yg} a formula for these matrices was derived.  

\section{Open string CFT}

Here a very brief introduction to open string conformal field
theory is given. Only those aspects needed in the rest of the paper
are mentioned. 

Open string conformal field theory is defined 
on surfaces with handles, boundaries and crosscaps. Any such
surface has a double cover which only has handles, and on which
one defines a closed, oriented conformal field theory. 
This CFT is the starting point for constructions of open (and 
unoriented) strings, which are referred to as ``open descendants" of
the closed string theories \cite{Sagnotti:1996eb}. 
The
presence of boundaries and crosscaps is described by 
boundary and crosscap ``states", which are not really states themselves,
but in fact non-normalizable linear combinations of states in
the closed string Hilbert space. 

The closed string CFT has a chiral algebra which includes the
Virasoro algebra. The boundary may preserve all or only part
of the closed string chiral algebra, but must at least 
preserve the Virasoro algebra.
Here we will assume that the entire chiral algebra
remains unbroken (``trivial gluing").
 The condition that 
a symmetry is not broken by a boundary or a crosscap
is
\be 
(J_n + (-1)^{h_J}\tilde J_n) \ket{B}=0 \ ;\ \ \ \ \ \ \
 (J_n + (-1)^{h_J+n} \tilde J_n) \ket{C}=0
\ee
where $J_n$ is a mode of a chiral current, $\tilde J_n$
a mode of an anti-chiral current, $\ket B$ a boundary state
and $\ket C$ a crosscap state. A basis for the solutions to
these conditions is formed by the Ishibashi states \cite{Onogi:1989qk}
\be \ket{B_i} = \sum_I \ket{I}_i \otimes U_B \ket{I}_{i^c} \ ;\ \ \ \ \ \ \ 
\ket{C_i}= \sum_I \ket{I}_i \otimes U_C \ket{I}_{i^c} \ .
\ee
Here the $i$ labels a representation of the chiral algebra and
$i^c$ its charge conjugate. The sum is over all states in the
representation, and $U_B$ and $U_C$ are operators satisfying
\be 
\tilde J_n U_B = (-1)^{h_J} U_B \tilde J_n\ ;\ \ \ \ \ \ \ 
\tilde J_n U_C = (-1)^{h_J+n} U_C \tilde J_n
\ee
Any boundary state must be a linear combination of
these Ishibashi states, {\it i.e.}
\be
\ket{B_a} = \sum_i B_{ai} \ket{B_i}\ ;\ \ \ \ \ \ \
\ket{C} = \sum_i \Gamma_{i} \ket{C_i} 
\ee 
It turns out that in general one can allow for several boundary
states, labelled by a boundary label $a$, but for only one
crosscap state. 

A choice of a set of boundary labels $a$, and a set of coefficients
$B_{ai}$ and $\Gamma_i$ form part of the data that define an
open string CFT. 
Although more is required to specify all correlation functions
on arbitrary surfaces, this information is sufficient 
to compute the one-loop diagrams without external lines that
contribute to the open and closed string partition functions. 
Hence we can at least compute the spectrum of the theory. 
The diagrams are computed in the transverse channel, in which
closed strings propagate between two boundaries, a boundary
and a crosscap, or two crosscaps. 
\begin{eqnarray}
&\hbox{Transverse Annulus:~~~~~~~~~~} &N_a N_b \bra{B^c_a}e^{i\tau H}\ket{B_b}
 \\ 
&\hbox{Transverse Moebius strip:~~~} &N_a \left[\bra{B^c_a}e^{i\tau H}
\ket{C}+\bra{C^c}e^{i\tau H}\ket{B_a}\right]\\ 
&\hbox{Transverse Klein bottle:~~~~~} &N_a \bra{C^c}e^{i\tau H}\ket{C} 
\end{eqnarray}
Here $H$ is the closed string Hamiltonian: 
$H=2\pi(L_0+\tilde L_0 - c/12)$, and $\tau$ is a real number
representing the length of the cylinder. The subscript ``$c$" indicates
that a CPT conjugate state is to be used.  
The integers $N_a$ are the
Chan-Paton multiplicities. 
One can express these amplitudes in terms of characters of the
representation $i$. 
By means of a transformation of the parameter $\tau$ one
can then compute the corresponding amplitudes in the direct channel
(the open and closed string loop channels). In the case of the
Klein bottle and the annulus this transformation acts on the characters
as the modular transformation matrix $S$, whereas in the case of
the Moebius strip one uses the matrix $P=\sqrt{T}ST^2S\sqrt{T}$, with
$\sqrt{T}$ defined as $\exp{i\pi(L_0-c/24})$. Then one arrives at
the following expressions 
\def\half{{\scriptstyle 1\over\scriptstyle 2}}
\begin{eqnarray}
&\hbox{Direct Annulus:~~~~~~~~~~}\hfill &\half N_a N_b A^i_{~ab} \chi_i(\half\tau) 
 \\ 
&\hbox{Direct Moebius strip:~~~}\hfill &\half N_a M^i_a \hat\chi_i(\half+\half \tau)\\ 
&\hbox{Direct Klein bottle:~~~~~}\hfill &\half K^i \chi_i(2\tau)  
\end{eqnarray}
Here $\hat\chi_i  \equiv T^{-\half}\chi_i $, and the parameter
$\tau$ is purely imaginary. The coefficients are
\be A^i_{~ab}= \sum_m S^i_{~m} B_{am} B_{am}\ ; \ \ \
M^i_{~a}= \sum_m P^i_{~m} B_{am} \Gamma_{m}\ ; \ \ \
K^i= \sum_m S^i_{~m} \Gamma_{m}\Gamma_{m}\ ; 
\ee

\subsection{Constraints}

These coefficients are subject to several constraints. Since the
direct channel contributions yield the open and closed
string partition functions, the coefficients are subject to the
requirement that all multiplicities should be positive integers (if one
applies the formalism to fermionic strings one would like to see
negative integers for space-time fermions, but those signs come out
automatically if one takes into account ghosts properly). This
yields two important conditions:\def\mod{{~\rm mod~}}
\begin{eqnarray}
&\hbox{Closed sector:~~~~~~}\hfill &| K^i | = Z_{ii}   
 \\ 
&\hbox{Open sector:~~~~~~~}\hfill & M^i_a  = A^i_{aa} \mod 2\ ,\ \ \
 | M^i_a | \leq A^i_{aa}\ \   
\end{eqnarray}
In writing down the first condition we assume that $Z_{ii} \leq 1$. One
can write down modular invariant partition functions for which that
is not the case, but this always means that some fields must be 
resolved into irreducible components first. Then the combination
$\half(Z_{ii}+K_i)$ is either a symmetric or anti-symmetric projection,
or vanishes completely.
In the second condition it
is assumed that $A^i_{~aa}$ is non-negative. Then 
$\half(N_a N_a A^i_{~aa}+N_a M^i_a)$ can be interpreted as
$\half(A^i_{~aa}+M^i_a)$ symmetric tensors of the Chan-Paton group,
plus $\half(A^i_{~aa}-M^i_a)$ anti-symmetric tensors. 

There are other constraints that are easy to check. First of all 
the ``completeness conditions" \cite{completeness} \begin{eqnarray}
 A_{ia}^{~~b}A_{jb}^{~~c}&=N_{ij}^{~~k} A_{ka}^{~~c} \\
  A_{iab}A^{i}_{cd}&=
A_{iac} A^{i}_{bd} 
\end{eqnarray}
Here the boundary indices are raised and lowered with the ``boundary metric"
$A^0_{~ab}=\sum_n S_{0n} B_{na}B_{nb}$, which must be order-2 permutation. In particular the matrix $A^0_{~ab}$ must have
 entries $0$ or $1$, and must be numerically equal to its own inverse $A^{0ab}$. The three allowed
Chan-Paton gauge groups correspond precisely to the
three allowed combinations of $A$ and $M$: $A^0_{aa}=M^0_a=1$ gives
$Sp(N_a)$ (if $N_a$ is even), $A^0_{aa}=-M^0_a=1$ gives
$SO(N_a)$, $A^0_{ab}=A^0_{ba}=1, M^0_a=M^0_b=0, a\not=b$ gives
$U(N_a)$ (if $N_a=N_b$).

It is convenient to define the matrices
\be R_{ia}={B_{ia} \sqrt{S_{i0}} }\ee
A sufficient condition for the completeness conditions as well as
the properties of $A^0_{ab}$ is then \cite{stanev}\cite{Behrend:1999bn}
$$ R_{ia}R_{ib}^*=\delta_{ab}  $$
$$ R_{ib}R_{jb}^*=\delta_{ij}    $$
This may not be the most general solution, but the only one 
considered here. Given the matrices $R$ it is natural to define also
\be U_{i}={\Gamma_{i}\sqrt{S_{i0}}}\ . \ee

Another easy constraint is the ``Klein bottle constraint" \cite{descendants}
\be
K_i K_j K_k \geq 0\ \ \hbox{~if~} N_{ijk} \not=0 
\ee
This ensures that the Klein bottle defines a consistent truncation on the 
spectrum. If this were not satisfied two states that are in the projected
spectrum can have couplings to a third state that is not in the projected
spectrum, so that the latter can appear in the intermediate channels of
tree-diagrams. A slightly weaker form of this
constraint was conjectured in \cite{Huiszoon:1999jw}.

There are other constraints (see \cite{sewing}), but 
they involve additional quantities
(such as OPE coefficients and duality matrices),  that are not readily
available, except in a few special cases.  
We will see, however, that the constraints described above are already
very restrictive. 

\section{Open strings and simple currents}

In this section various simple current modifications of the
basic open descendant construction are discussed. This basic
construction is often referred to as the ``Cardy case", and consists
of a natural ansatz for the boundaries, supplemented with an ansatz
for the crosscap. We will discuss this case first, and show that 
the crosscap ansatz follows directly from the boundary ansatz and
the positivity requirements.

\subsection{Uniqueness of crosscaps in the Cardy case}

Cardy \cite{Cardy:1989ir} conjectured a general ansatz for the annulus
$$ B_{ai} = {S_{ai}\over \sqrt{S_{0i}}} $$
This ansatz satisfies the completeness condition, and yields an annulus 
coefficient equal to the Verlinde fusion coefficients for $S$. It follows
that all fields propagate in the transverse channel, and hence that
all fields $\phi_{i,i^c}$ must appear in the bulk theory. Therefore
this ansatz
requires the torus partition function to be defined in terms of $Z=C$,
the charge conjugation matrix.

To check any of the other conditions we need to know the crosscap
coefficients $\Gamma_i$. They were first determined in an $SU(2)$
model by Sagnotti et. al. and this result was used as a conjecture
for all other cases. Additional support for this conjecture was given
in \cite{Huiszoon:1999xq}, where it was shown that this conjecture 
satisfies all positivity and integrality constraints; see also 
\cite{Borisov:1998nc}\cite{Gannon:1999is}
for related results and 
\cite{Bantay:1997zk}\cite{Bantay:2000xb}
for a discussion of the Klein bottle constraint.

The results of the latter paper can in fact be turned around: we may
impose positivity and integrality and derive the crosscap coefficients.
To do this we make use of the fact that the set of boundaries in the
Cardy case is in one-to-one correspondence 
with the bulk labels, and in particular there is a boundary ``0". Then
$A^i_{00}=N^i_{00}=\delta_{i0}$, and hence $M^i_0=\pm 1$.
Hence
\be
\sum_m P^i_{~m} B_{0m} \Gamma_{m}= \pm \delta_{i0} 
\ee
from which we read off immediately
\be
 \Gamma_{m}= \pm {P_{0m}\over \sqrt{S_{0m}}} \ ,
\ee
where the sign is undetermined, but does not depend on $m$ (it would
ultimately determined by the tadpole cancellation condition).
We conclude that the crosscap coefficients are in fact uniquely determined
by the boundary coefficients (up to an overall sign, which is not fixed
by any CFT constraint).  
Recently the crosscap coefficient has also been determined using 
3-dimensional TFT \cite{Felder:1999mq}, but the argument given here 
has the advantage of being considerably simpler.
It does not necessarily
generalize to other annuli, because the boundary ``0" need not exist.
But it will generalize to the cases considered below.

\subsection{Simple current modifications}

Various simple current related modifications of the Cardy ansatz
have been studied. A possibly incomplete list is 

\begin{itemize}
\item{Extensions of the closed chiral algebra}
\item{Non-trivial Klein bottle projections }
\item{Simple current automorphisms } 
\item{Broken bulk symmetries} 
\end{itemize}

The first item is dealt with entirely within the closed string theory
and requires no further discussion. Examples of the second kind have
been around for a while \cite{descendants}\cite{properties}, and were studied in general in \cite{Huiszoon:1999xq} where also the consistency
conditions were shown to hold in general. In the third class the
bulk theory is defined by means of a simple current automorphism of
the fusion rules. Such examples were first studied in 
\cite{nondiagonal}. In \cite{Fuchs:1997kt} automorphisms of spin-$\half$
simple currents were studied in general. These authors gave an 
interpretation of the boundary label ``$a$" in terms of the label of
representations of a ``classifying algebra", which in the C-diagonal
case is just the Verlinde algebra. A remarkable feature of this case is
the appearance of the fixed point resolution matrix of the spin-$\half$
current in the formulas for the boundary coefficient. This matrix does not
play a direct r\^ole in the closed string theory.  The fourth case is
studied for example in \cite{Fuchs:1998fu}\cite{Recknagel:1998sb}\ and \cite{Fuchs:1999zi}\cite{Fuchs:1999xx}. The latter papers 
reveal an even more interesting
appearance of fixed point resolution matrices. I will not discuss any
of these results in detail here, but in the rest of this section I will
show how in the second and third case one may also derive the crosscap
coefficients directly from the positivity constraints. 

\subsection{A formula for $P$}

To derive these results I will need a formula relating matrix elements
of the matrix $P$ on simple current orbits. It is analogous to the
well-known formula for $S$ \cite{Intriligator:1990zw}\cite{ScYb}
\be
  S_{a,J^{\ell} b}=
e^{2\pi i \ell Q_J(a)}S_{ab}  
\ee
The corresponding relation for $P$ is more complicated due
to the factors $\sqrt{T}$ in the definition of $P$, and only
works if the indices are shifted by even powers of the current 
\be \label{eq:Prel}
 P_{a,J^{2\ell} b}=\rho(\ell)e^{2\pi i \Delta(2l,b)}
e^{2\pi i Q_J(a)}P_{ab} 
\ee
where
\be
 \Delta(\ell,c)=h_{J^{\ell}c}-h_{J^{\ell}}-h_c + {\ell} Q(c) 
\ee
with $0 \leq Q < 1 $,  and
\be
\rho(\ell)=e^{\pi i (r\ell +M_{2\ell})}
\ee
where
\be
 M_{\ell}=h_{J^{\ell}}-{r{\ell}(N-{\ell})\over 2N}\ ,  
\ee
where $r$ is the monodromy parameter of the current.
The derivation is straightforward, provided one replaces the ill-defined
quantity $\sqrt{T}$ systematically by the well-defined quantity 
$\exp(i\pi(h-c/24))$. The first two factors in (\ref{eq:Prel}) are signs. 

\subsection{Non-trivial Klein bottle projection}

The first simple current modification I will consider was called
a ``non-trivial Klein bottle projection" in \cite{Huiszoon:1999xq}, because
in some cases it produces sign changes in the coefficients $K_i$
with respect to the Cardy case. Here I will study it from a different
starting point, namely the annulus.

Consider the following set of reflection coefficients,
 $R_{ma}=S_{ma} \sqrt{{S_{m0}\over S_{mJ}}}$, which satisfy the
completeness conditions.  
Then
$$A^i_{~00}= \sum_m {S^i_{~m} R_{ma} R_{mb} \over S_{mJ}}=N^{Ji}_{~~00}$$
Hence $M^i_{~0}=\pm \delta^{Ji}_0=\pm \delta^{i}_{J^c}$.
On the other hand
$$ M^i_{~0}=\sum_m P^i_{~m}  U_m\sqrt{{S_{m0}\over S_{mJ} }} $$
so that
\begin{eqnarray}
 U_m &= \sum_i P_{mi}M^{i}_{~0}\sqrt{S_{mJ}\over S_{m0} } \\
&= \pm P_{mJ^c} \sqrt{S_{mJ}\over S_{m0}} \\
\end{eqnarray}

Now we use the formula
$$ P_{a,K^2c}=\epsilon(K,c)e^{2\pi i Q_K(a)}P_{ac} $$
where $\epsilon(K,c)$ is a sign. We choose $K=J^c$ and $c=J$. Then
\begin{eqnarray}
U_m&=\pm P_{mJ^c} \sqrt{S_{mJ}\over S_{m0}}\\
&=\pm e^{2\pi i Q_{J^c}(m)}P_{mJ}
\sqrt{S_{mJ}\over S_{m0}}\\
&=\pm P_{mJ}
\sqrt{S_{m0}\over S_{mJ}}
\end{eqnarray}
This is the formula used in \cite{Huiszoon:1999xq}, which turns out 
to be the only possibility, given the annulus coefficients. The
proof that the other constraints are satisfied can be found in that paper.

\subsection{Non-trivial simple current automorphism}

Up to now the bulk modular invariant was the charge conjugation invariant,
$Z=C$. A general modular invariant  is characterized by a  left and
right extension
of the chiral algebra, the modules of which are paired by an automorphism.
For the construction of ``open descendants" the left  and right extensions
must be identical, and the automorphism symmetric. The extension  can be
dealt with at the closed string level, which leaves the possibility of
non-trivial automorphisms. One may distinguish three basic types: 
the charge conjugation invariant, simple current invariants and 
anything else, which by definition is ``exceptional". 

A solution is known for simple current automorphisms generated by
$\Zbf_{\rm odd}$ and $\Zbf_2$ simple currents. 
Both cases are described by 
the following formula for $R$
$$ R_{m,a_i}={1\over \sqrt {| G |}} \sum_{J \in S_a}
\sum_{K \in G/S_a} {\breve S}^J_{m,Ka} \psi^J_i $$
Here $G$ is the full simple current group that produces
the automorphism, $S_a \subset G$ is the stabilizer of $a$,
${\breve S}^J$ is the (appropriate generalization of) the orbit
Lie algebra $S$-matrix corresponding to $J$ and $\psi^J_i$ is
a discrete group character of $S_a$. The boundary labels are
taken to correspond to be $G$-orbits, with each orbit 
split into $|S_a|$ components. The label $m$ ranges over all fields
with $Z_{mm^c}=1$.
This formula is in any case
correct for $G=\Zbf_2$ (in which case it summarizes the four 
expressions given in \cite{Fuchs:1997kt})
and $G=\Zbf_{\rm odd}$, and is a good
candidate for a general formula, although this has not been investigated
yet. From here on we will assume that $G=\Zbf_N$,
 with $N$ odd or equal to 2.

In \cite{Huiszoon:1999jw}\ a consistent ansatz was presented for
the crosscap coefficient. Here I will show how to derive it, demonstrating
that this ansatz is in fact unique (given the reflection coefficients).
Consider the ``zero-boundary" $a=0$.
It is not hard to show that
$$ A^i_{~00}=\sum_J \delta^i_J\ , $$
so that
$$ M^i_{~0}=\eta(i) \sum_J \delta^i_J\ , $$
where $\eta(i)$ is a sign.

Computing $M^i_{~0}$ directly gives
\be
\label{eq:ssum}
 M^i_{~0}=\sum_{m, Q(m)=0} {P^i_{~m} \sqrt{| G |} U_m }\ .  
\ee
The restriction to zero charge is due to the fact that $R_{m0}$ vanishes
if $m$ has non-zero charge. Using the inverse of $P$ we get,
for $Q(m)=0$
\be
\label{eq:sssum}
 U_m = {1\over \sqrt{| G |}}\sum_{J\in G} \eta(J)P_{Jm}\ , 
\ee
Note that we get no information about the coefficients
for non-zero charge. On the other hand,
the fact that the terms with  $Q(m)\not=0$ do not contribute
to the sum (\ref{eq:ssum}) implies that  
$$ \sum_{J\in G} \eta(J)P_{Jm} = 0 \hbox{~~~for~} Q(m)\not= 0 $$
This puts strong constraints on the possible choices for 
$\eta(J)$.  

We can write (\ref{eq:sssum}) as a sum over even and odd
elements:
$$ V_m = {1\over \sqrt{N}}\left[
\sum_{\ell=0} \eta(2\ell )P_{J^{2\ell},m} 
+\sum_{\ell=0} \eta(2\ell+1 )P_{J^{2\ell+1},m}\right]
$$
Here $V_m=U_m$ if $Q(m)=0$ and $V_m=0$ (with $U_m$ undetermined)
if $Q(m)\not=0$.
Using the  relation (\ref{eq:Prel}) we can express the first terms
in terms of $P_{0m}$ and the second ones in terms of $P_{Jm}$. If
$N$ is odd we can 
furthermore 
express  $P_{Jm}=P_{J^{N+1}m}$ in terms of $P_{0m}$. Then the
final result can be expressed as
$$ V_m = {1\over \sqrt{N}}P_{0m}  
\sum_{\ell=0}^{N-1} \sigma(\ell)e^{2\pi i l Q(m)} \ ,
 $$
where $\sigma(\ell)$ is a sign. 
This can only vanish for all $Q(m)\not=0$ if $\sigma(\ell)=\pm 1$, 
independent of $\ell$. Then
also $U_m=V_m$ for $Q(m)=0$ is determined,
\be \label{eq:umdef} U_m= \pm {\sqrt{N}}P_{0m} 
 \ee

For $N=2$ we find 
$$ V_m = {1\over \sqrt{2}}\left[\eta(0) P_{0m}+ \eta(J) P_{Jm}\right] \ .$$
Since $P_{0m}$ and  $P_{Jm}$ are unrelated we cannot simplify the result
further. In addition $P_{0m}=P_{Jm}=0$ for $Q(m)=\half$, so that we get
no further constraints. It may seem that there are two solutions now
(not counting the overall sign), but closer inspection of the positivity
constraints of fixed point boundary labels reveals that for each CFT
only one definite relative sign $\eta(0)/\eta(J)$ is
 allowed. 
The other sign is the correct one for a different choice of
reflection coefficients  (see \cite{Huiszoon:1999jw}\ for more details). 

To determine the coefficients $U_{m}$ for charged $m$ we 
may use closed sector positivity for bulk label $i=0$. This
leads to the requirement $\sum_m U_m^2=1$.
For $N$ odd there are no transverse
channel fields with non-zero charge, and it is easy to show that
(\ref{eq:umdef}) satisfies this condition.
In the case $N=2$ there is an additional problem, namely that we do
not know $U_m$ for fixed points. Allowing for an unknown contribution
on the fixed points we get
$$ K^i = \sum_{m, Q(m)=0} {S^i_{~m} U_m U_m \over S_{0m}}+
\sum_{f, Jf=f} {S^i_{~f} U_f U_f \over S_{0f}}=K^i_1+K^i_2\ , $$
where $U_m=V_m$ given above, and $U_f$ is unknown. It can be
shown \cite{Huiszoon:1999jw}
that $K_1$ already satisfies the positivity constraints. Then
the only allowed values for $K_2$ are $K_2^i=k_i K_1^i$, where
$k_i=-2$ or $0$. But it is easy to show that 
$K_2^{Ji}=-K_2^i$ whereas $K_1^{Ji}=K_1^i$. Then we must have 
$K_2^i=0$ and hence $U_f=0$. So also in this case the crosscap
is unique.

\subsection{Other automorphisms}

For automorphisms that are not simple current modifications of
the $C$-invariant there is an amusing observation to be made. The
coefficients $U_m$ must vanish whenever $m$ is not paired with
its charge conjugate. The coefficients $K^i$ must vanish if $i$ is not
paired with itself. This leads to the sum rule
$$ \sum_{i, Z_{ii}=1} K^i S_{im} = 0\ , \hbox{if~} Z_{mm^c}=0 $$
For example in the interesting case $Z={\bf 1}$ this leads to the sum rule
$\sum_i K^i S_{im}=0$ for complex fields $m$. Empirically this rule
is indeed satisfied, with all $K^i$ equal to 1. Although in some
cases (e.g $A_2$ level 1)
this sum rule is satisfied in a trivial way, there are many other
examples (e.g $A_2$ level 3) where it is non-trivial, and implies
relations between matrix elements of $S$ that are hard to derive in
any other way. This illustrates that by studying open, unoriented CFT one
may learn something interesting about the closed, oriented case.

\vspace{10mm}

\begin{center}
{\bf Acknowledgements}
\end{center}

I would like to thank Lennaert Huiszoon and Nuno Sousa for discussions and
Christoph Schweigert for comments on the manuscript. 
Special thanks to the organizers for inviting me to give this talk,
and for making this a successful conference despite what turned out
to be very difficult circumstances.

\vspace{10mm}


\end{document}